\documentclass[twocolumn,showpacs,amsmath,amssymb,aps]{revtex4}

\usepackage{epsfig}
\usepackage{graphicx}
\usepackage{dcolumn}
\usepackage{bm}

\begin{document}

\preprint{preprint}

\title{Electronic structure of ternary palladates and effect of hole doping: A valence band photoemission spectroscopic study}

\author{B. H. Reddy, Asif Ali and Ravi Shankar Singh}
\email{rssingh@iiserb.ac.in}
\affiliation{Department of Physics, Indian Institute of Science Education and Research Bhopal, Bhopal Bypass Road, Bhauri, Bhopal - 462 066, INDIA}

\date{\today}

\begin{abstract}

We investigate the electronic structure of ternary palladates $A$Pd$_3$O$_4$ ($A$ = Sr, Ca) using valence band photoemission spectroscopy and band structure calculations. Overall width of the valence band and energy positions of various features in experimental valence band spectra are well captured by band structure calculations using hybrid functional. Band structure calculations within local density approximations lead to metallic ground state while the calculations using hybrid functional provide band gap of 0.25 eV and 0.22 eV for CaPd$_3$O$_4$ and SrPd$_3$O$_4$ respectively, suggesting moderated to strong electron correlation strength in these narrow band gap semiconducting palladates. High resolution spectra reveals negligibly small intensity at Fermi level, $E_F$, for parent compounds while hole doped SrPd$_3$O$_4$ (by 15\% Li substitution at Sr site) exhibits a Fermi cut-off suggesting metallic character in contrast to semiconducting transport. These observations reveal the importance of localization of electrons in case where the Fermi edge falls in the mobility edge.

\end{abstract}


\maketitle

Insulator to metal transition (IMT) in transition metal oxides (TMOs) has been extensively studied in past few decades \cite{MIT}. Strongly correlated 3$d$ TMOs have attracted much attention due the fact that large electron correlation ($U$) plays crucial role in deciding the exotic ground states in these systems \cite{Mott}. On the other hand, among 5$d$ TMOs where electron correlation is expected to be significantly weak but the relativistic effects comes to play and observation of spin orbit coupling (SOC) driven Mott insulating state is realised in some Ir based oxides \cite{bjkim}. 4$d$ TMOs are in the intermediate regime where SOC as well as $U$ are expected to be moderately small compared to 5$d$ and 3$d$ systems respectively. Owing to these, Ruthenates \cite{ruthenates} and Rhodates \cite{rhodates} have been extensively studied. The next member of 4$d$ family is Pd which acquires the PdO$_4$ square planer geometry in most of the palladium based oxides. While going from octahedral configuration to square planer configuration, the degeneracy of $e_g$ orbital lifts up and $d_{z^2}$ orbital goes lower in energy, leaving $d_{x^2-y^2}$ orbital completely empty in Pd 4$d^8$ configuration. This leads to insulating state in most of the palladium based oxides. A finite density of states due to the completely filled $d_{z^2}$ band just below $E_F$ in these systems allows hole doping driven IMT which has been observed in hole doped PdO \cite{pdo}, PbPdO$_2$ \cite{pbpdo} and many other palladium oxide systems.

Narrow band gap semiconducting ternary palladates have attracted significant attention in recent past due to observation of electron/hole doped IMT \cite{APO-MIT,APO-Na} and for the possibility of being an excitonic insulator \cite{APO-ExIn}. A recent theoretical investigation suggests Dirac semi-metallic ground state having three sets of Weyl nodes in the vicinity of Fermi edge in these palladium oxides \cite{APO-Weyl}. In this paper, we investigate the electronic structure of CaPd$_3$O$_4$, SrPd$_3$O$_4$ and hole doped SrPd$_3$O$_4$ (by 15\% Li substitution at Sr site) using photoemission spectroscopy and band structure calculations to understand the role of electron correlation and hole doping on the IMT in these systems.

The samples were prepared by a conventional solid state reaction method in the polycrystalline form using high purity ingredients (CaCO$_3$, SrCO$_3$, Li$_{2}$CO$_3$ and PdO powders). Well ground mixtures were palletized and sintered at a 700~$^o$C for more than 3 days with three intermittent grinding and palletization. All the samples were furnace-cooled. The $x$-ray diffraction patterns collected at room temperature exhibit single phase samples without any signature of impurity. Cubic lattice constants are found to be 5.74~\AA, 5.83~\AA~ and 5.82~\AA~for CaPd$_3$O$_4$, SrPd$_3$O$_4$ and Sr$_{0.85}$Li$_{0.15}$Pd$_3$O$_4$ respectively which are in well agreement with earlier reports \cite{APO-MIT,APO-Na,no-ExIn}. Temperature dependent $x$-ray diffraction and Raman spectroscopic experiments were also performed and will be reported elsewhere \cite{surajit}. Electrical resistivity measurements are also in well agreement with literature exhibiting insulating nature of all the samples while Sr$_{0.85}$Li$_{0.15}$Pd$_3$O$_4$ exhibits reduced resistivity than SrPd$_3$O$_4$ \cite{APO-MIT,APO-Na,no-ExIn}.

Photoemission measurements were performed using a Scienta R4000 electron energy analyzer at a base pressure of $\sim$ 4$\times$10$^{-11}$~torr. The experiments were performed using various photon sources and the total instrumental resolutions were set to 0.4~eV for Al~$K\alpha$ (1486.6~eV), 8~meV for He~{\scriptsize II} (40.8~eV) and 5~meV for He~{\scriptsize I} (21.2~eV) radiation (energy). Clean sample surfaces were achieved by {\em in-situ} fracturing and the cleanliness was ascertained by tracking the sharpness of O~1$s$ and absence of C~1$s$ core levels. The band structure calculations were performed using full-potential linearized augmented plane-wave (FPLAPW) method as implemented in Wien2k \cite{Wien2k} for experimentally found lattice parameters. 8$\times$8$\times$8 $k$ mesh within the first Brillouin zone and Hybrid functional (YS-PBE0) \cite{YS-PBE0} were used to calculate the density of states (DOS). The criteria for energy and charge convergence were set to 0.1 meV and 10$^{-4}$ electronic charge per formula unit (fu) respectively.

\begin{figure}[tb]
\centering
\includegraphics[width=1.1\textwidth,natwidth=1200,natheight=750]{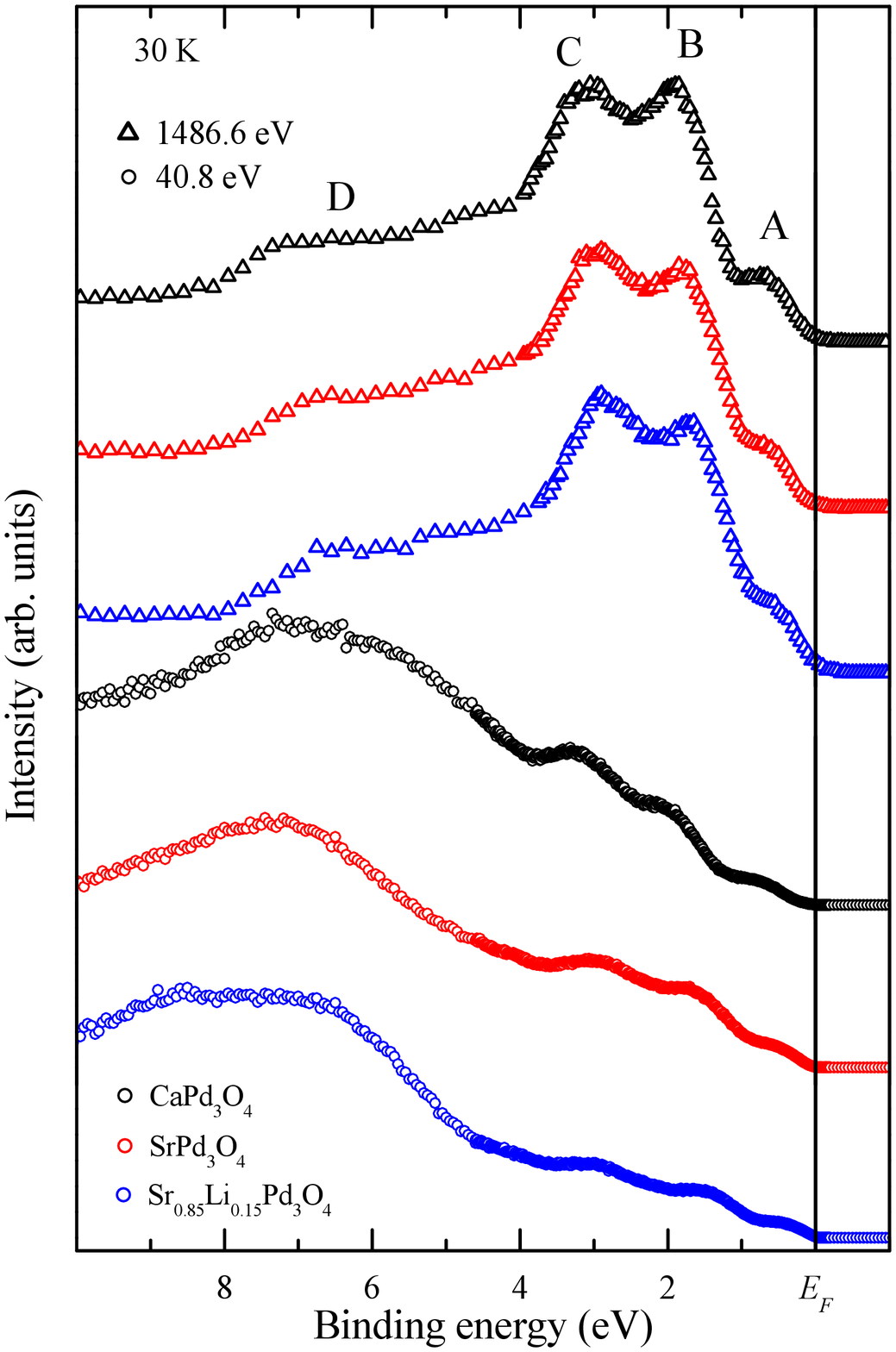}
\vspace{-15ex}
\caption{\label{fig:epsart} (color online) Valence band photoemission spectra at 30 K using Al~$K\alpha$ (open triangles) and He~{\scriptsize II} (open circles) radiations for CaPd$_3$O$_4$ (black), SrPd$_3$O$_4$ (red) and Sr$_{0.85}$Li$_{0.15}$Pd$_3$O$_4$ (blue).}
\end{figure}

\begin{figure}[tb]
	\centering
	\includegraphics[width=1.1\textwidth,natwidth=1200,natheight=750]{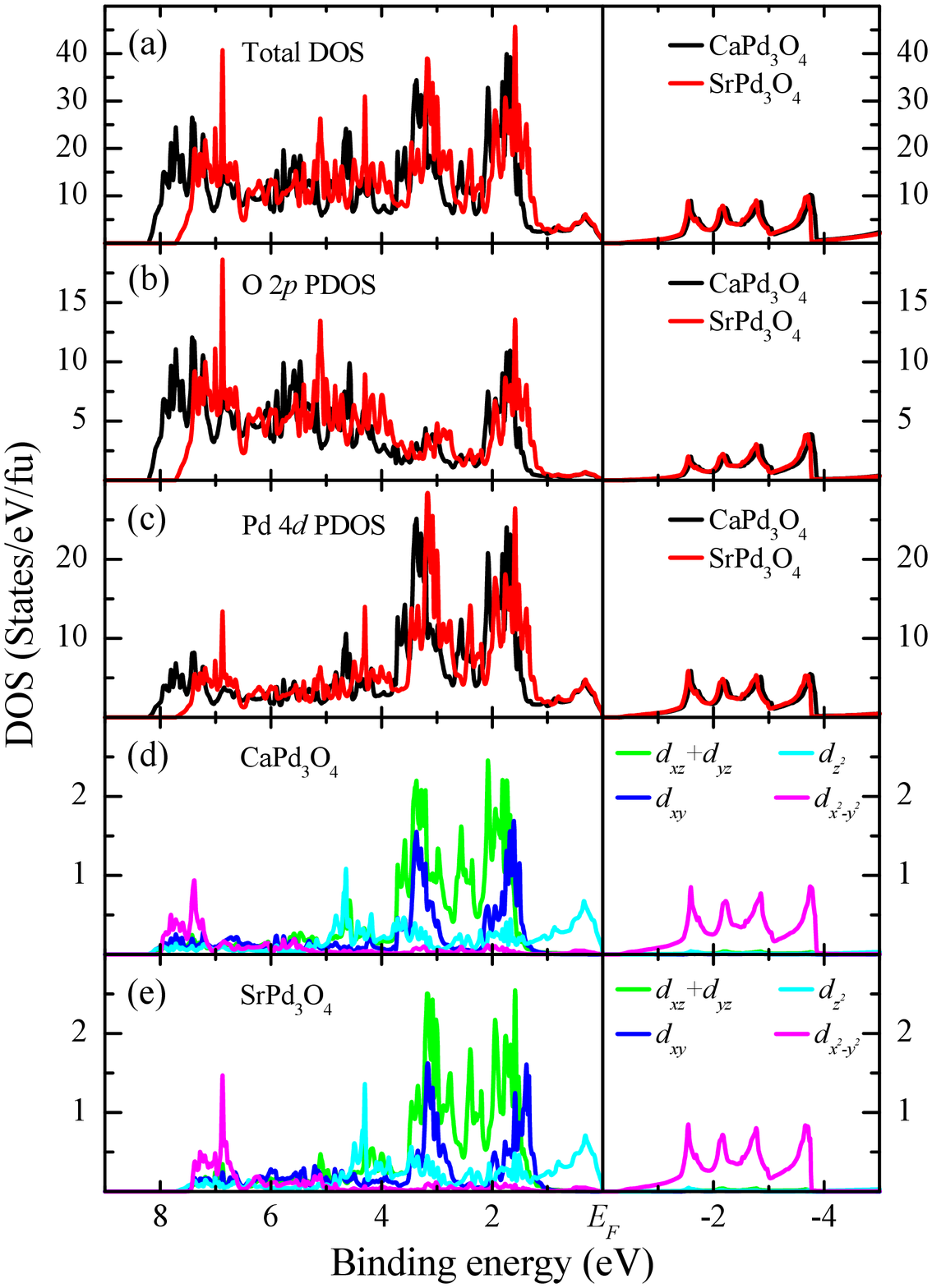}
	\vspace{-15ex}
	\caption{\label{fig:epsart} (color online) (a) Total DOS (b) O 2$p$ partial DOS (c) Pd 4$d$ partial DOS in CaPd$_3$O$_4$ and SrPd$_3$O$_4$. $d_{xz}$+$d_{yz}$, $d_{xy}$, $d_{z^2}$ and $d_{x^2-y^2}$ band contributions are shown in (d) for CaPd$_3$O$_4$ and in (e) for SrPd$_3$O$_4$.}
\end{figure}

Fig. 1. shows the valence band photoemission spectra of APd$_3$O$_4$ [A = Ca, Sr, Sr$_{0.85}$Li$_{0.15}$] collected at 30 K using monochromatic Al~$K\alpha$ radiation (open triangles) and monochromatic He~{\scriptsize II} radiation (open circles) as excitation sources. All the spectra are normalized to the highest intensity and the valence band exhibits four prominent features with peak like structures A, B and C appearing at $\sim$ 0.5 eV, $\sim$ 2 eV, $\sim$ 3 eV binding energy respectively and a broad feature D appearing between 4 eV to 9 eV binding energy. The valence band in these systems are formed by hybridization of O 2$p$ and Pd 4$d$ bands. Since the photo-ionization cross-section of Pd 4$d$ states relative to O 2$p$ states is much larger in case of Al~$K\alpha$ spectra with respect to He~{\scriptsize II} spectra \cite{yeh}, the features A, B, and C can be attributed to Pd 4$d$ derived states and broad feature D can be attributed to O 2$p$ derived states. 

To understand these results, we show the total density of states (TDOS) in Fig. 2(a) as well as O 2$p$ and Pd 4$d$ partial density of states (PDOS) in the Fig. 2(b) and Fig. 2(c) respectively for CaPd$_3$O$_4$ and SrPd$_3$O$_4$. Similar to the experimental spectra there are four sets of feature appearing in the occupied energy range. A broad feature appears in $\sim$ 4 eV to 8 eV binding energy range having predominantly O 2$p$ character followed by two peak like structures at $\sim$ 3 eV and $\sim$ 2 eV binding energy along with the forth feature appearing below 1 eV binding energy corresponding to features D, C, B and A respectively of the experimental spectrum as shown in the Fig. 1. In the experimental spectra, features B and C shift about 0.1 eV towards lower binding energy while going from CaPd$_3$O$_4$ to SrPd$_3$O$_4$ which can also be seen clearly in the calculated results. This may due to the larger lattice constant and thus larger Pd-O distance leading to smaller crystal field splitting in case of SrPd$_3$O$_4$ (Pd-O bond lengths of 2.058 \AA) in comparison to CaPd$_3$O$_4$ (Pd-O bond lengths of 2.029 \AA). A further shift of about 0.1 eV of the whole spectra towards lower binding energy while going from SrPd$_3$O$_4$ to Sr$_{0.85}$Li$_{0.15}$Pd$_3$O$_4$ is due to the hole doping in the system which leads to shift of the $E_F$ within the occupied states. The intensity at the $E_F$ is negligibly small in case of CaPd$_3$O$_4$ and SrPd$_3$O$_4$ while it is finite in case of Sr$_{0.85}$Li$_{0.15}$Pd$_3$O$_4$ suggesting that the Fermi edge falls at the band edge.

It is to note here that the band structure calculation within local density approximations (LDA) or generalized gradient approximations (GGA) do not produce the band gap while GGA+$U$ ($U$ = 2 eV) and calculations using TB-mBJ potentials leads to opening of small gap suggesting that the electron correlation plays an important role in these system \cite{AminKhan}. Band structure calculations using screened hybrid functional have been quite successful in capturing the band gaps in correlated electron systems where the ground states are often semiconducting and/or insulating \cite{YS-PBE0,YS-PBE0-1}. In the present case of CaPd$_3$O$_4$ and SrPd$_3$O$_4$, the observed band gaps of about 0.25 eV and 0.22 eV are very similar to those observe from transport measurements. Total width of the valence band and observed features along with the shift in them while going from CaPd$_3$O$_4$ to SrPd$_3$O$_4$ are also well captured in these hybrid calculations suggesting influence of electron correlation in these ternary palladates.

\begin{figure}[tb]
	\centering
	\includegraphics[width=1.1\textwidth,natwidth=1200,natheight=750]{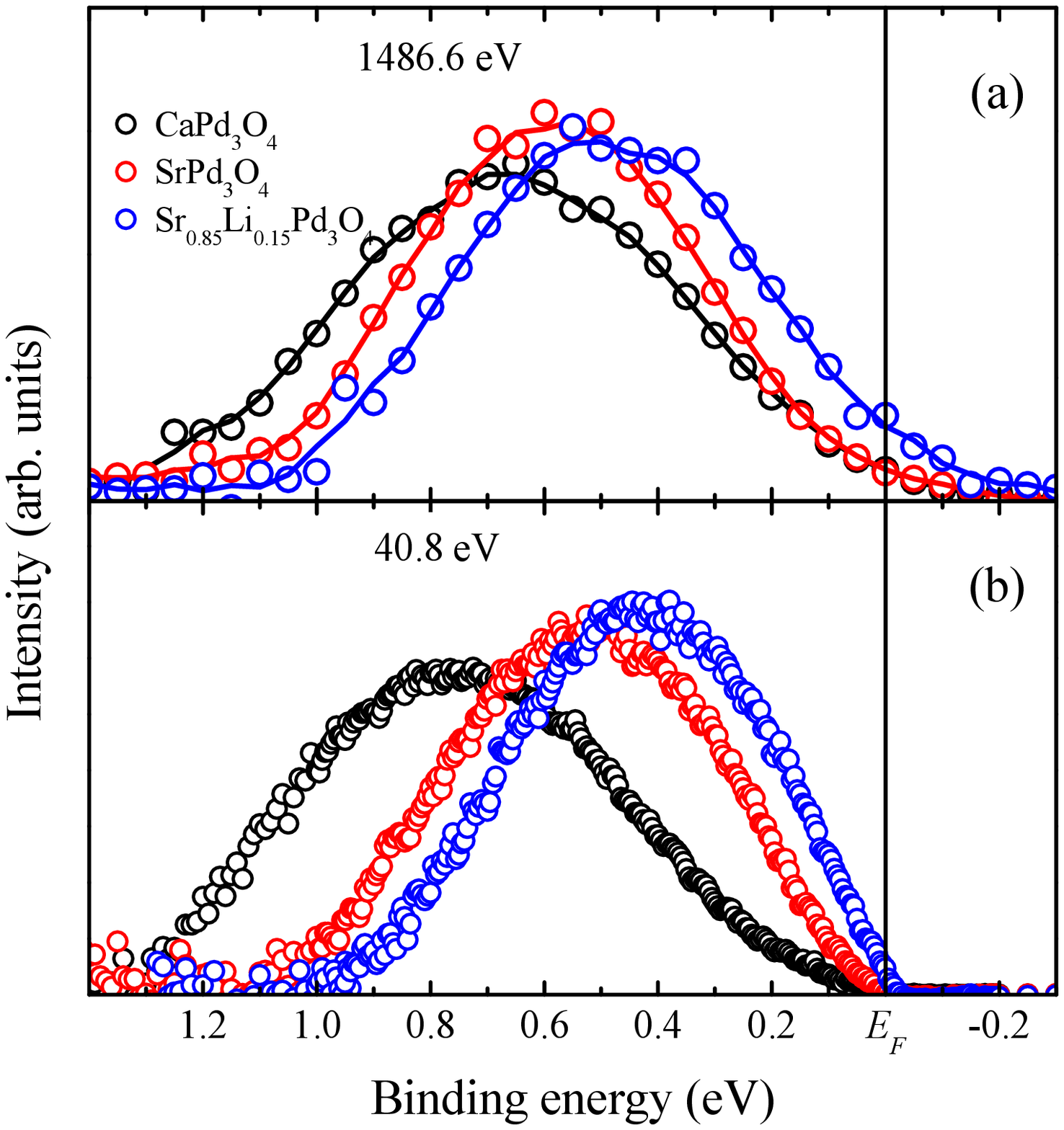}
	\vspace{-38ex}
	\caption{\label{fig:epsart} (color online) (a) Feature A extracted from Al~$K\alpha$ spectra for CaPd$_3$O$_4$ (black), SrPd$_3$O$_4$ (red) and Sr$_{0.85}$Li$_{0.15}$Pd$_3$O$_4$ (blue). Lines represent the guide to eye. (b) Feature A extracted from He~{\scriptsize II} spectra for CaPd$_3$O$_4$ (black), SrPd$_3$O$_4$ (red) and Sr$_{0.85}$Li$_{0.15}$Pd$_3$O$_4$ (blue).}
\end{figure}

\begin{figure}[tb]
	\centering
	\includegraphics[width=1.1\textwidth,natwidth=1200,natheight=750]{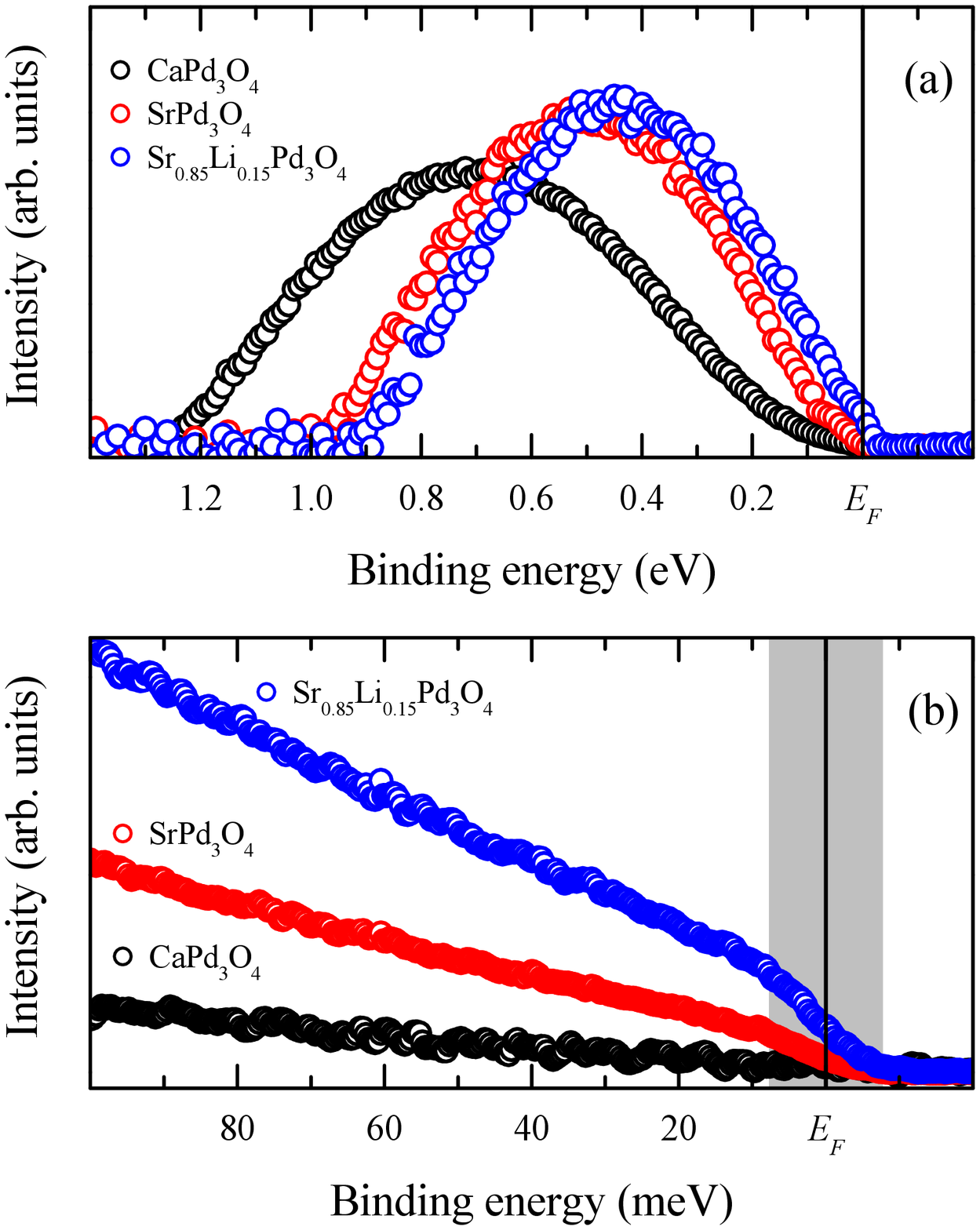}
	\vspace{-28ex}
	\caption{\label{fig:epsart} (color online) (a) Feature A extracted from He~{\scriptsize I} spectra for CaPd$_3$O$_4$ (black), SrPd$_3$O$_4$ (red) and Sr$_{0.85}$Li$_{0.15}$Pd$_3$O$_4$ (blue). (b) High resolution He~{\scriptsize I} spectra in the vicinity of $E_F$. Shaded region shows the $\pm ~ 3k_bT$ energy range at 30 K.}
\end{figure}

Fig. 2(d) and 2(e) shows various Pd 4$d$ band contributions for CaPd$_3$O$_4$ and SrPd$_3$O$_4$ respectively. Going from octahedral crystal field to square planer configuration, the degeneracy of the $e_g$ band lifts up and the $d_{z^2}$ band goes lower in energy. Additionally, Pd-Pd interaction between two PdO$_4$ square planer lying on top of each other redistributes the energies of these orbitals and the configuration becomes $d_{xz}$/$d_{yz}$, $d_{xy}$, $d_{z^2}$ and $d_{x^2-y^2}$ in the increasing order of energy \cite{ramsheshadri}. In the present case Pd 4$d$ band having 8 electrons leads to completely empty $d_{x^2-y^2}$ band. It is clear from the Fig. 2(d) and Fig. 2(e) that the feature A corresponds to the topmost filled $d_{z^2}$ band in both CaPd$_3$O$_4$ and SrPd$_3$O$_4$ with negligible O 2$p$ band contributions. Thus we extract the near $E_F$ feature A by delineating the spectra by fitting it with multiple Gaussians. The extracted spectra has been shown in Fig. 3(a) and Fig. 3(b) corresponding to Al~$K\alpha$ spectra and He~{\scriptsize II} spectra respectively for all three systems. All the spectra are normalized by the total integrated intensity. Al~$K\alpha$ spectra exhibit peaks at $\sim$ 0.7 eV, $\sim$ 0.6 eV and $\sim$ 0.5 eV binding energy for CaPd$_3$O$_4$, SrPd$_3$O$_4$ and Sr$_{0.85}$Li$_{0.15}$Pd$_3$O$_4$ respectively as shown by symbols in Fig. 3(a) and lines represent guide to the eye. The spectral trend and shift in the peak positions are very similar for He~{\scriptsize II} spectra which suggests that the photon energy does not have much influence on the spectra. These two spectra collected at two different photon energies provide different probing depth in this technique and the similarity with respect to peak position and width of the band suggests that the surface and bulk electronic structure are very similar in these systems. The major difference appears in the close vicinity of $E_F$ which appears presumably due to the higher resolution employed in case of He~{\scriptsize II} spectra. The spectral intensity in the vicinity of $E_F$ is much smaller for CaPd$_3$O$_4$ in comparison to SrPd$_3$O$_4$ and spectral intensity becomes zero at the $E_F$ for both the systems. These results suggest that both the systems are insulating while CaPd$_3$O$_4$ having larger resistivity as seen in transport measurements may be attributed to smaller DOS in the vicinity of $E_F$. Interestingly Sr$_{0.85}$Li$_{0.15}$Pd$_3$O$_4$ exhibits finite intensity at $E_F$ indicating metallic ground state.

To understand the results further, we investigate the near $E_F$ spectra using high intensity He~{\scriptsize I} radiation which also provides better energy resolution of about 5 meV. We show the feature A, extracted and normalized by the integral intensity for He~{\scriptsize I} spectra for all three systems in the Fig. 4(a). All the spectra exhibit very similar lineshapes and peak positions when compared with He~{\scriptsize II} spectra shown in Fig. 3(b). The near $E_F$ region has been expanded and shown in the Fig. 4(b). The shaded grey region shows the 3$k_BT$ energy range corresponding to 30 K temperature. Spectral intensity for CaPd$_3$O$_4$ and SrPd$_3$O$_4$ show monotonous decrease and becomes zero at $E_F$, while CaPd$_3$O$_4$ having much smaller intensity in the vicinity of $E_F$. This observation is commensurate with the higher resistivity of CaPd$_3$O$_4$ than SrPd$_3$O$_4$. Interestingly Sr$_{0.85}$Li$_{0.15}$Pd$_3$O$_4$ shows even higher intensity in the vicinity of $E_F$ and a clear Fermi cut-off is observed as seen within the shaded region. The observation of Fermi edge suggests metallic character in contrast to insulating/ semiconducting transport. It is interesting to note here that the hole doped ternary palladates $A$Pd$_3$O$_4$ exhibit insulator to metal transition for more that 20\% Li/Na substitution at the place of $A$ site \cite{no-ExIn,APO-MIT,APO-Na}. Observed Fermi cut-off for Sr$_{0.85}$Li$_{0.15}$Pd$_3$O$_4$ with insulating transport suggests that the electrons at the $E_F$ being at the band edge with low Fermi velocity are possibly getting localised in the presence of intrinsic disorder which leads to Anderson type insulating phase in this system \cite{anderson}.

In conclusion, we report the valence band photoemission spectroscopic study of the ternary palladates and hole doped systems. Band structure calculation in conjunction with photoemission spectra suggests moderate to strong electron correlation strength. The parent systems show zero intensity at $E_F$ suggesting insulating behaviour. Spectral intensity around $E_F$ corresponding to CaPd$_3$O$_4$ is much smaller than that of SrPd$_3$O$_4$ which is commensurate with transport measurements, where higher resistivity is observed for CaPd$_3$O$_4$. Hole doping in SrPd$_3$O$_4$ leads to rigid band shift of about 0.1 eV in turn leading to higher spectral intensity in the vicinity of $E_F$. Sr$_{0.85}$Li$_{0.15}$Pd$_3$O$_4$ exhibits a clear Fermi cut-off in the high resolution spectra corresponding to metallic ground state in contrast to insulating transport suggesting importance of disorder in this system which leads to localization of electrons at $E_F$ which falls in the mobility edge. 

Authors acknowledge the support of Central Instrumentation Facility and HPC Facility at IISER Bhopal. Support from DST-FIST (Project No. SR/FST/PSI-
195/2014(C)) is also thankfully acknowledged.

\end{document}